# Cyberbullying or just Sarcasm? Unmasking Coordinated Networks on Reddit

Pinky Pamecha[1,2], Chaitya Shah[1,3], Divyam Jain[1,4], Kashish Gandhi[1,5], Kiran Bhowmick[1,6]

and Meera Narvekar[1,7]

[1]*Department of Computer Engineering, Dwarkadas J. Sanghvi College of Engineering*
*Mumbai, India*

[2]*thepinkypamecha@gmail.com*, [3]*chaitya0623@gmail.com*, [4]*jain.divyam.271@gmail.com*, [5]*kashishgandhi6112003@gmail.com*,
[6]*kiran.bhowmick@djsce.ac.in*, [7]*meera.narvekar@djsce.ac.in*

*Abstract*— With the rapid growth of social media usage, a common trend has emerged where users often make sarcastic comments on posts. While sarcasm can sometimes be harmless, it can blur the line with cyberbullying, especially when used in negative or harmful contexts. This growing issue has been exacerbated by the anonymity and vast reach of the internet, making cyberbullying a significant concern on platforms like Reddit. Our research focuses on distinguishing cyberbullying from sarcasm, particularly where online language nuances make it difficult to discern harmful intent. This study proposes a framework using natural language processing (NLP) and machine learning to differentiate between the two, addressing the limitations of traditional sentiment analysis in detecting nuanced behaviors. By analyzing a custom dataset scraped from Reddit, we achieved a 95.15% accuracy in distinguishing harmful content from sarcasm. Our findings also reveal that teenagers and minority groups are particularly vulnerable to cyberbullying. Additionally, our research uncovers coordinated graphs of groups involved in cyberbullying, identifying common patterns in their behavior. This research contributes to improving detection capabilities for safer online communities.

## I.　INTRODUCTION

Social media serves as a platform that provides individuals with an opportunity to connect, share, and engage globally. The rise of social media platforms has transformed the way people interact, fostering a virtual space where ideas and expressions can be shared openly and instantaneously [1]. While social media offers many benefits, it also presents drawbacks, having given rise to new forms of negative behavior, such as cyberbullying.

Recently, cyberbullying has emerged as one of the major issues on social media. Studies have shown a marked increase in cyberbullying incidents correlating with the growing usage of social media [2], particularly among younger demographics who are the most active on these platforms. Cyberbullying is defined as "an aggressive act or behavior that is carried out using electronic means by a group or an individual repeatedly and over time against a victim who cannot easily defend him or herself" [3]. This behavior can adversely affect a person's mental health, leading to social anxiety, depression, stress, and social isolation [4]. Research has shown that people in minority groups [5] are more vulnerable to cyberbullying attacks, and individuals from different cultural backgrounds may perceive textual context differently, leading to more confusion and personal attacks as arguments escalate [6].

The anonymity and wide reach offered by social media have exacerbated the issue, enabling perpetrators to target individuals with ease and often without immediate consequences. Approximately 46% of teenagers in America experience cyberbullying [7]. This form of bullying has both physical and mental impacts on the victims [8]. In extreme cases, victims resort to self-destructive actions, such as suicide, as a result of the trauma of cyberbullying, which can be unbearable [9]. Therefore, the identification and prevention of cyberbullying are crucial.

One of the major problems in identifying and addressing cyberbullying lies in the complex nature of online communication. Sarcasm and humor, common elements in digital interactions, can often mask harmful behavior. While sarcasm is generally used to express irony or satire, in the context of cyberbullying, it can disguise harmful comments, insults, and threats. This ambiguity creates a significant challenge for detection and moderation systems, which may find it difficult to accurately interpret the true intent behind comments that appear harmless to the system.

One way cyberbullying is masked is through the use of sarcasm, where harmful intent is hidden behind seemingly humorous or ironic remarks. For instance, a comment like, "Wow, I'm so impressed by how you manage to fail at everything you try, it's a real talent!" may appear lighthearted or joking on the surface, but its underlying message is intended to belittle and demean the recipient. This type of sarcastic remark allows the bully to downplay their intentions, making it easier to dismiss the comment as a joke if confronted. The victim, however, is left feeling insulted and isolated, while the bully benefits from the plausible deniability provided by the sarcastic tone. This subtlety makes it particularly challenging for others, including detection systems and bystanders, to identify the comment as harmful.

Traditional methods of detection, which rely heavily on sentiment analysis, may not sufficiently differentiate between a sarcastic joke and an insult. This challenge highlights the necessity for more advanced methodologies that can effectively distinguish between harmless and harmful sarcasm. Additionally, cyberbullying often involves groups of people who coordinate their actions to target specific victims. These groups can be particularly difficult to identify, as they create continuous and coordinated efforts to harm others. Detecting and understanding the dynamics of these groups is essential for developing effective prevention measures against cyberbullying.

This research seeks to address these challenges by developing a framework to differentiate cyberbullying from sarcasm on social media platforms. By employing natural language processing and machine learning techniques, our paper aims to distinguish between cyberbullying and sarcasm. Additionally, the research focuses on identifying and analyzing groups involved in cyberbullying, uncovering their coordination and strategies. The ultimate goal is to enhance the detection capabilities of social media platforms, thereby contributing to safer and more respectful online communities.

## II. BACKGROUND

The increasing use of social media has not only revolutionized communication but has also brought about significant challenges, including the rise of cyberbullying. As online interactions grow, so does the prevalence of harmful behavior, making cyberbullying detection an important area of study. Researchers have been keen to explore various approaches to mitigate this issue, primarily focusing on the use of machine learning and natural language processing (NLP) techniques to analyze the vast amounts of textual data generated on social media platforms [10].

Existing literature shows that a wide range of methods, from basic sentiment analysis [11] to more advanced machine learning models, have been employed to identify instances of cyberbullying. For example, some studies have utilized traditional machine learning models like Support Vector Machines (SVM) and Naïve Bayes (NB) [12] to classify cyberbullying content. While these methods have provided some success, they often fall short when dealing with the more subtle and nuanced forms of online abuse, such as those involving sarcasm or context-dependent language.

A growing body of research suggests that deeper insights into user interactions and the complex nature of online conversations are essential for more accurate detection. Social media platforms like Twitter, Reddit, and Instagram are filled with both direct insults and less obvious forms of aggression, such as sarcasm, which can make hurtful comments seem innocuous. This complexity has led researchers to explore more sophisticated models that can better understand context and intent, such as deep learning approaches [13]. These models aim to go beyond simply identifying harmful words, seeking instead to grasp the underlying meaning and tone of interactions.

One significant challenge identified in the literature is the difficulty of detecting sarcasm, which often serves as a mask for cyberbullying. Sarcasm detection requires models that can understand not just the words used but also the implied meanings that can be contradictory to the literal words [14]. This requires advanced models capable of contextual understanding—a capability that many earlier systems lacked [15].

Despite advancements, there remains a noticeable gap in research where sarcasm and cyberbullying detection are integrated into a single framework. Existing models tend to handle these aspects separately, often leading to misclassifications where sarcastic remarks are either ignored or incorrectly flagged. The integration of these detection capabilities is crucial, as it would allow for a more nuanced understanding of online interactions, distinguishing between genuine, benign communication and harmful content disguised as humor or sarcasm.

Additionally, the literature highlights the under-explored area of coordinated group dynamics in cyberbullying. Most studies have focused on individual cases without delving into how groups of users might work together to target individuals. Understanding these dynamics is critical, as coordinated attacks can amplify the impact of bullying and make it more pervasive.

This research seeks to address these gaps by developing a comprehensive approach that not only differentiates between sarcasm and cyberbullying but also analyzes the networks of individuals involved in such activities. By mapping out these networks, the study aims to provide a clearer picture of how cyberbullying operates on a broader scale and to develop strategies that can effectively disrupt these harmful behaviors.

## III. METHODOLOGY

In this study, a multi-step methodology was implemented to detect coordinated cyberbullying on Reddit, focusing on the distinction between sarcasm and cyberbullying. First, Researchers compiled a set of predefined keywords, including offensive and cuss words, which served as the basis for scraping relevant posts from Reddit. These posts were then processed through a sarcasm detection model to filter content that might be intended as humor or irony rather than direct harassment. Posts identified as sarcastic or non-sarcastic were further analyzed for cyberbullying using a specialized classification model.

This approach allowed examination of linguistic differences between sarcasm and cyberbullying, isolating posts where harmful intent was present. To identify coordinated behavior, network analysis was performed on the users and posts, seeking patterns of interaction, shared language, and collective targeting of individuals. By combining keyword analysis, sarcasm detection, and cyberbullying classification with network analysis, a comprehensive framework was developed for understanding how cyberbullying is orchestrated on Reddit.

### A. Data Collection

The initial keyword list used in this study was sourced from a compilation developed by researchers at Carnegie Mellon University [16], specifically by Luis von Ahn's Research Group. Luis von Ahn's research focuses on creating systems that leverage the combined strengths of humans and computers to address large-scale problems that neither can solve independently. While this list includes certain words that may not be universally perceived as offensive, it provides a solid foundation for those seeking to identify offensive or profane language in online environments. This serves as a valuable starting point to detect offensive and profane terms.

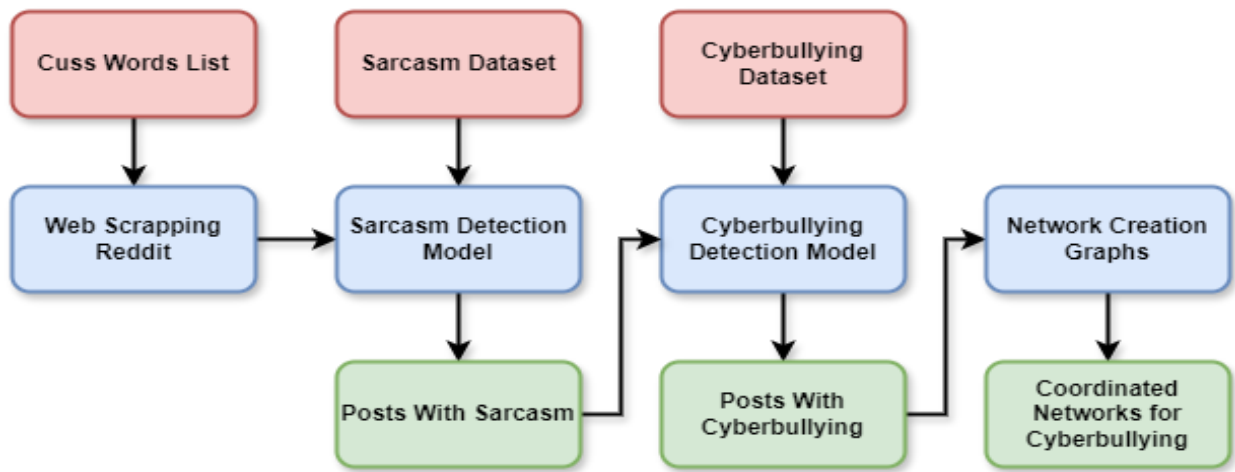

*Fig. 1 Model Architecture*

- **Sarcasm Dataset**

Past studies in Sarcasm Detection were on twitter datasets collected using hashtags. Such datasets were very noisy in terms of labels and language. Also, in order to detect sarcasm in replies, one needs to also consider the previous tweets to fully get contextual awareness. To overcome this, the News Headlines dataset for Sarcasm Detection was used [17]. This dataset is collected from two news websites: TheOnion and HuffPost. TheOnion helps us get sarcastic versions of current events, while real and non-sarcastic news headlines are collected from HuffPost. This dataset contained 28,619 rows which were classified as sarcastic or not sarcastic, thereby helping us provide a comprehensive basis for the model to learn and identify sarcastic language effectively.

- **Cyberbullying Dataset**

To train the cyberbullying detection model, a merged dataset was utilized comprising 113,915 rows from three distinct sources, all obtained from Kaggle. The Cyberbullying Tweets dataset provided a balanced sample, with an equal distribution of cyberbullying and non-cyberbullying instances, ensuring the model could accurately distinguish between harmful and non-harmful content. The Cyberbullying Classification [18] dataset contributed 47,000 tweets categorized into various forms of cyberbullying, including age, ethnicity, gender, religion, other types of bullying, and non-cyberbullying, with around 8,000 entries per class. Additionally, the Cyber Bullying Hinglish dataset added 17,068 new rows, further enhancing our ability to identify cyberbullying.

- **Testing Dataset**

In this research, Reddit was selected as the primary source for scraping comments and data due to its large, diverse, and active user base. This made it an ideal platform, providing a rich dataset for analysis. Reddit's openness and the availability of vast amounts of data facilitated thorough collection and exploration of textual content, which was essential for developing and testing our detection models. This approach not only deepened our understanding of the interplay between cyberbullying and sarcasm but also enabled the mapping of networks engaged in these harmful activities, providing a more comprehensive analysis of the issue.

Initially, an attempt was made to our model using a pre-existing Twitter dataset from a prior study. However, significant challenges were encountered when trying to gather fresh Twitter data using web scraping techniques and cyberbullying-related keywords. Twitter's recent rate limits, capping us at 600 posts per session, hindered our data collection efforts. Alternatives such as Nitter instances were also explored but were proved inadequate due to missing or incomplete data.

Subsequently, Reddit was chosen as our primary data source, employing keyword searches and web scraping tools. To collect relevant data from Reddit, we employed a multi-step scraping process using both Selenium and BeautifulSoup. Initially, we utilized Selenium to navigate Reddit and extract post-level information based on predefined keywords. Our scraping approach leveraged the flexibility of web drivers, allowing for dynamic interactions with web elements that are difficult to access through static scraping methods.

We began by using Selenium's *find_elements* function to locate the posts based on their HTML structure. Specifically, we targeted the following XPATH: *//div/reddit-feed/faceplate-tracker*, which corresponds to the main container of posts on Reddit. For each identified post, we extracted key attributes, such as the post's tracking data (*data-faceplate-tracking-context*), source, and various post-specific metadata including post ID, author ID, score, number of comments, and URL.

In this process, JSON data embedded in each post's HTML was parsed to extract information such as the post's author ID, creation timestamp, and whether the post was marked as NSFW or archived. This allowed us to retrieve detailed data points for each post, which were then stored for further analysis. These posts were tagged with the keyword that led to their discovery, enabling us to trace the content back to specific topics or language patterns related to cyberbullying or sarcasm.

Once the post URLs were collected, the second phase of scraping involved visiting each post's page to extract the associated comments. For this, we switched to BeautifulSoup to parse the HTML structure of the comments section. Using *find_all*, we targeted the *shreddit-comment* tag, which contains individual comments.

Using BeautifulSoup, we extracted information for each comment, including the author, depth (which indicates whether the comment is a reply to another), parent ID, and score. The *Paragraph Content* field stored the actual textual content of the comment, which we extracted from the *div* tag with the ID *-post-rtjson-content*. This allowed us to capture the entire thread of comments for each post, facilitating a deeper analysis of interactions between users, especially in cases of coordinated cyberbullying.

To further investigate coordinated cyberbullying, approximately 860,000 rows of data were scraped from Reddit, encompassing a wide range of posts and comments across various subreddits. This dataset facilitated the differentiation between sarcasm used as a disguise for cyberbullying and genuine non-harmful content. Additionally, it allowed for an analysis of network dynamics, aiding in the detection of coordinated bullying efforts that were concealed under the guise of sarcasm.

### B. Implementation

The model selection process began with the testing of baseline models, such as Logistic Regression, which achieved an accuracy of 86.77%. Additionally, experiments were conducted with Random Forest and Support Vector Machine (SVM) classifiers, both of which yielded satisfactory results. However, these traditional models lacked the depth and sophistication required to effectively detect cyberbullying, particularly given the complex and nuanced nature of online language, including sarcasm [19]. Acknowledging these limitations, BERT (Bidirectional Encoder Representations from Transformers) and RoBERTa (Robustly Optimized BERT Pre-Training Approach) were trained on the dataset to more accurately detect both cyberbullying and sarcasm.

Both BERT and RoBERTa have the ability to analyze text in a bidirectional manner, meaning they consider the context of both preceding and succeeding words. This makes them highly effective for understanding complete sentences and accurately identifying emotions such as hate speech, threats, or sarcasm, which is critical for this study. These models have been pre-trained on large datasets of diverse language patterns and structures. This pretraining allows them to generalize better when applied to our cyberbullying dataset, reducing the risk of overfitting and ensuring more accurate predictions. BERT and RoBERTa have consistently demonstrated state-of-the-art performance on language-related tasks, including sentiment analysis and offensive language detection. When applied to the dataset, they significantly outperformed traditional models, showcasing their ability to capture subtle linguistic nuances, making them ideal for detecting both sarcasm and cyberbullying. BERT was chosen for this task due to its demonstrated superior performance compared to other models, achieving the highest accuracy on our dataset .To detect cyberbullying and sarcasm, both tasks leverage BERT (Bidirectional Encoder Representations from Transformers), a powerful transformer-based model that captures nuanced language features, such as context, sentiment, and word relationships.

To implement the BERT base uncased model, a structured process was followed. Before training, the text data was converted into numerical data through BERT tokenization. This process transforms raw text into tokens that the model can interpret, ensuring the sequences are of manageable lengths (truncated to 128 tokens when necessary). The BertForSequenceClassification is pre-trained on a general corpus and fine-tuned for the specific task of cyberbullying detection. The model has two output labels: "Cyberbullying" and "Not Cyberbullying". The transformer layers of BERT help capture the contextual information from the input sequence, allowing it to identify aggressive, harmful, or bullying behavior. The training process involves passing tokenized input sequences to BERT and using the cross-entropy loss to update model weights. The model is optimized with Adam optimizer, and it learns to classify texts as cyberbullying or not based on labeled training data.

TensorDataset was employed to split the dataset into training and testing sets. For the training dataset, we used the RandomSampler to ensure randomized batches, while the SequentialSampler was used for the testing dataset, maintaining the order of data for evaluation. The model was trained using the Adam optimizer for 5 epochs.This optimization technique is well-suited for complex models like BERT, helping to adjust the learning rate dynamically during training to achieve better performance.

### IV. RESULTS

After training, the BERT model achieved an impressive accuracy of 95.15%, making it the most suitable model for our research. The model's ability to detect nuanced language and subtle distinctions between sarcasm and cyberbullying was particularly effective in this study. To evaluate the model's performance, we used metrics such as accuracy score and the classification report, which provided detailed insights into precision, recall, and F1-score. These metrics confirmed the robustness and reliability of BERT in detecting coordinated cyberbullying efforts on Reddit, making it the ideal solution for our use case.

Table I. Result Metrics.

| Metric | Value |
| --- | --- |
| Mean Absolute Error | 1846.39 |
| Mean Squared Error | 13367089.49 |
| Root Mean Squared Error | 3656.10 |
| R-Squared | 0.95 |

Table II. Different Model Accuracies.

| Model | Accuracy(in percentage) |
| --- | --- |
| Logistic Regression | 86.77 |
| SVM | 88.14 |
| Random Forest Classifier | 90.74 |
| RoBERTa | 94.71 |
| BERT | 95.15 |

The network graph illustrates the relationships between authors and Subreddits on Reddit, focusing on detecting instances of cyberbullying masked as sarcasm. In this visualization, green nodes represent specific subreddits topics, while pink nodes correspond to individual authors. Authors are clustered around subreddits topics based on their posts, providing a visual representation of engagement within specific discussions.

The size of each green node correlates with the number of posts associated with that subreddit.The size of each pink node represents the number of posts made by the author on a particular topic. Authors with correspondingly larger nodes, suggests a higher level of involvement or dominance in the conversation. This can help in identifying key contributors or influential figures in cyberbullying or sarcastic exchanges.

Based on the analysis of Reddit's top five subreddits that engage in both cyberbullying and sarcasm, distinct patterns were identified that aid in differentiating sarcasm from cyberbullying, although some subreddits exhibit overlapping behaviors:

**Subreddits Exhibiting Both Cyberbullying and Sarcasm**: Subreddits like *shitposting* and *Tiktokcringe* are often known for ironic humor and pushing boundaries of what's considered acceptable. The overlap of cyberbullying and sarcasm in these spaces suggests that users frequently mix humor with harmful intent, which can easily go unnoticed by moderation systems. from results, *teenagers* is a more vulnerable group that frequently engages in sarcastic banter, but the behavior can escalate into cyberbullying, as seen in these network patterns.

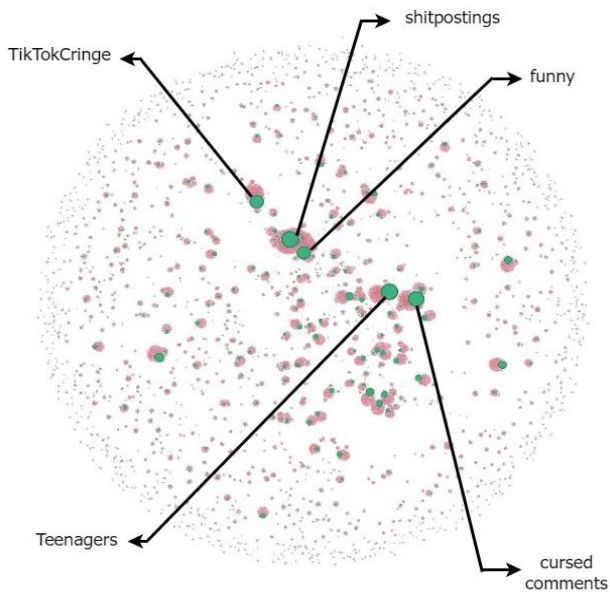

*Fig.2. Network graph for instances where both cyberbullying and sarcasm detected*

**Subreddits Exhibiting Only Sarcasm**: Subreddits such as *funny* and *Memes*, sarcasm appears to be used primarily for entertainment, without the intent to harm. These spaces foster a culture where irony and satire are tools for humor rather than aggression. Interestingly, *Teenagers* shows up again, but in this context, it suggests that the users here employ sarcasm for social bonding rather than cyberbullying.Sarcasm in *HistoryMemes* and *worldnews*, for example, is often employed to comment on historical or current events, often in a playful or critical tone without crossing into harassment territory.

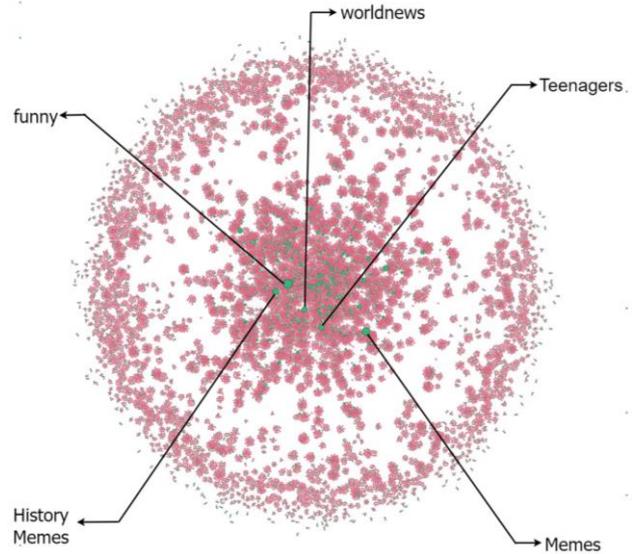

*Fig.3. Network graph for instances where only sarcasm detected*

**Subreddits Exhibiting Only Cyberbullying**: On subreddits such as *BlackPeopleTwitter* and *unpopularopinion* we found clear pattern of direct, aggressive behavior. These communities often engage in heated debates, and without sarcasm to mask intent, the bullying is explicit and targeted. *PoliticalHumor* and *Gaming* subreddits also emerge as spaces where users may directly attack others over differing political views or gaming-related disputes.*Teenagers* appears again, reinforcing that younger users are frequent targets or perpetrators of cyberbullying, often within environments that should be more carefully moderated.

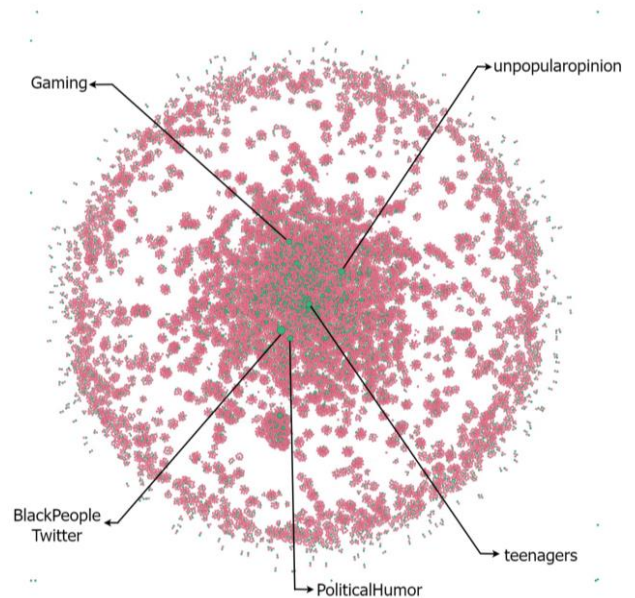

*Fig.4. Network graph for instances where only cyberbullying detected*

## CONCLUSION

This research contributes to the understanding and detection of coordinated cyberbullying on Reddit by distinguishing between sarcasm and cyberbullying, two often overlapping forms of online communication. By leveraging a keyword-based scraping methodology and utilizing advanced sarcasm detection models, , the study effectively filtered content for further analysis. The findings highlight that while sarcasm can be a common rhetorical device used humorously, it can also serve as a vehicle for harmful behavior when used in conjunction with cyberbullying. This distinction is crucial for the development of robust detection models aimed at mitigating online harassment.

Furthermore, the research uncovered the existence of coordinated networks of individuals engaged in cyberbullying. These groups utilize specific patterns, often cloaked in sarcasm or indirect language, to target individuals or communities. Identifying these coordinated efforts opens doors for more precise intervention strategies, including real-time detection and disruption of toxic group behavior. Future work can enhance this framework by incorporating more sophisticated linguistic models and network analysis techniques to further refine the distinction between humor and harm, and to uncover more nuanced forms of coordinated cyberbullying.

## FUTURE SCOPE

While the analysis of coordinated networks provides valuable insights, it is important to acknowledge the limitations of the study and explore potential areas for improvement.

- Approximately 860,000 rows of data were collected by analyzing only 167 out of 1,383 keywords. For each keyword, posts were filtered by selecting only those with more than 20 comments to manage computational constraints and focus on posts likely to reveal a coordinated network. Consequently, not all Reddit posts related to these keywords were included in the analysis. Expanding the dataset to incorporate the posts that were initially excluded could potentially yield more comprehensive and insightful results.
- Additionally, the analysis is currently limited to Reddit. Expanding the scope to include a cross-platform analysis would provide a more comprehensive understanding of targeted cyberbullying groups, offering deeper insights into both the topics discussed and the types of users being targeted across different platforms.